\documentclass[11pt, a4paper]{article}
\usepackage{tikz}
\usepackage{float}
\usepackage{amsfonts}
\usepackage{jinstpub}

\usepackage[ruled,vlined]{algorithm2e}
\usetikzlibrary{decorations.pathreplacing}
\usetikzlibrary{shapes.misc, positioning}

\newcounter{lastnote}

\newcommand\blfootnote[1]{%
  \begingroup
  \renewcommand\thefootnote{}\footnote{#1}%
  \addtocounter{footnote}{-1}%
  \endgroup
}

\definecolor{iocolor}{RGB}{169,205,236}
\definecolor{hlvcolor}{RGB}{224,235,245}
\definecolor{hscolor}{RGB}{166,166,166}
\definecolor{layercolor}{RGB}{218,218,218}
\definecolor{featcolor}{RGB}{202,203,229}
\definecolor{grucolor}{RGB}{222, 236, 239}

\definecolor{hlvlinecolor}{RGB}{0, 148, 197}
\definecolor{featurelinecolor}{RGB}{0, 43, 152}
\definecolor{hslinecolor}{RGB}{85, 85, 85}

\title{Anomalous Jet Identification via Sequence Modeling} 

\author[a]{Alan Kahn,}
\author[a]{Julia Gonski,}
\author[b]{In\^{e}s Ochoa,}
\author[a]{Daniel Williams,}
\author[a]{and Gustaaf Brooijmans}

\affiliation[a]{Nevis Laboratories, Columbia University, Irvington, NY, USA}
\affiliation[b]{Laboratory of Instrumentation and Experimental Particle Physics, Lisbon, Portugal}

\emailAdd{alan.kahn@cern.ch}
\emailAdd{julia.gonski@cern.ch}
\emailAdd{ines.ochoa@cern.ch}
\emailAdd{dmw2174@columbia.edu}
\emailAdd{gustaaf.brooijmans@cern.ch}

\abstract{This paper presents a novel method of searching for boosted hadronically
  decaying objects by treating them as anomalous elements of
  a contaminated dataset.  
 A \textit{Variational Recurrent Neural Network} (VRNN) is used to model jets as sequences of constituent four-vectors. 
After applying a pre-processing method which boosts each jet to the same reference mass, energy, and orientation, the VRNN provides each jet an \textit{Anomaly Score} that distinguishes between the structure of signal and background jets.
The model is trained in an entirely unsupervised setting and without high level variables, making the score more robust against mass and $p_{T}$ correlations when compared to methods based primarily on jet substructure. 
Performance is evaluated on the jet level, as well as in an analysis context by searching for a heavy resonance with a final state of two boosted jets. 
The Anomaly Score shows consistent performance along a wide range of signal contamination amounts, for both two and three-pronged jet substructure hypotheses. 
Analysis results demonstrate that the use of Anomaly Score as a classifier enhances signal sensitivity while retaining a smoothly falling background jet mass distribution. 
The model's discriminatory performance resulting from an unsupervised training scenario opens up the possibility to train directly on data without a pre-defined signal hypothesis.}

\begin{document}

\pagenumbering{gobble}

\baselineskip16pt

\maketitle 

\setlength{\abovedisplayskip}{5pt}
\setlength{\belowdisplayskip}{5pt}
\setlength{\abovedisplayshortskip}{0pt}
\setlength{\belowdisplayshortskip}{0pt}





\pagenumbering{arabic}


\section{Introduction}

Many of the open questions in physics, such as the nature of dark matter or the origin of the universe's matter-antimatter asymmetry, could potentially be addressed by the ATLAS and CMS experiments at the CERN Large Hadron Collider (LHC).\blfootnote{A code package demonstrating the model used in this study can be found at: \url{https://github.com/akahn1213/vrnn}.}
Since the discovery of the Higgs boson in 2012~\cite{atlas_higgs,cms_higgs}, no signals of new beyond the Standard Model (BSM) physics have been found. 
However, sophisticated analysis tools utilizing novel machine learning models hold promise to increase the sensitivity to rare signals. 
 
While many machine learning applications at the LHC focus on classifying established signatures, such as top quarks or Higgs bosons, there are a number of emerging techniques aimed at the discovery of new particles. 
Many of these techniques use classifier models to search for particular model hypotheses, by training and evaluating a model using Monte Carlo simulation of the BSM signal.
While these models show promising performance, they are developed to identify a specific model hypothesis, and often have little to no sensitivity to other models. In addition, they are subject to inaccuracies between simulated samples and data, such as those which occur in the non-perturbative regime of Quantum Chromodynamics (QCD) processes.

One way to circumvent these issues is to develop a model that trains directly on data, without requiring the need for simulated inputs.
Distinguishing new physics from Standard Model background at the LHC without a signal hypothesis is a novel and promising effort~\cite{Heimel_2019, deepAutoencoders, CWoLa, cheng2021variational, bortolato2021bump}. 
This paper demonstrates a way to identify BSM signals that present as anomalous substructure in jets using unsupervised, data-driven anomaly detection. 

Anomaly detection refers to a process in which anomalous elements are identified within a dataset that is mostly homogenous, but contaminated with outliers. 
In a machine learning context, this can be done with a model that learns an underlying distribution of data points, as characterized by high-level features of the data. 
Such a model can then identify out-of-distribution data solely on how poorly they are represented by the learned underlying distribution. 
Several candidate architectures have been developed for this purpose, and are described in the following subsections.
The examination of their features is instructive in choosing an architecture for the task of anomalous jet identification.

\subsection{Autoencoders}

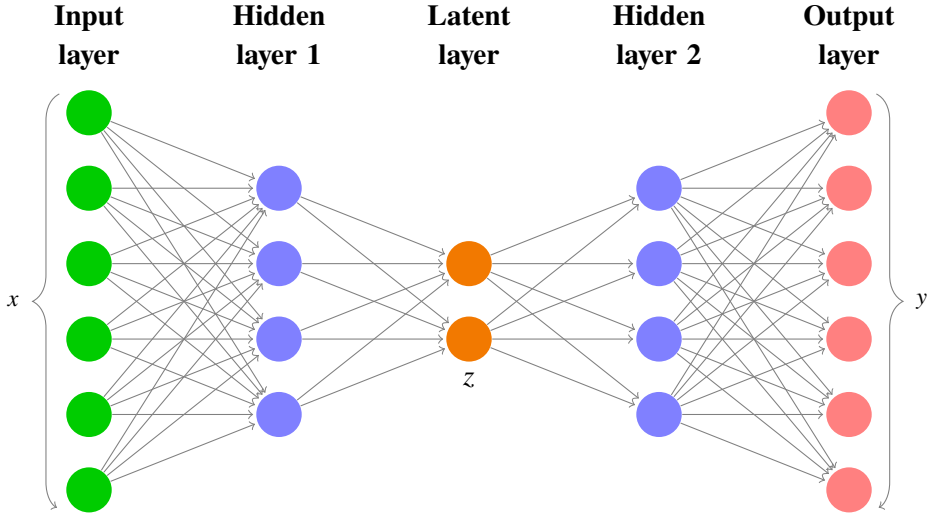
\begin{figure}[H]
  \begin{center}
  
    \def\layersep{2.5cm}
    
    \begin{tikzpicture}[shorten >=1pt,->,draw=black!50, node distance=\layersep]
        \tikzstyle{every pin edge}=[<-,shorten <=1pt]
        \tikzstyle{neuron}=[circle,fill=black!25,minimum size=17pt,inner sep=0pt]
        \tikzstyle{input neuron}=[neuron, fill=black!20!green];
        \tikzstyle{output neuron}=[neuron, fill=red!50];
        \tikzstyle{latent neuron}=[neuron, fill=black!5!orange];
        \tikzstyle{hidden neuron}=[neuron, fill=blue!50];
        \tikzstyle{annot} = [text width=4em, text centered]
    
        \foreach \name / \y in {1,...,6}
            \node[input neuron] (I-\name) at (0,-\y) {};
    
        \foreach \name / \y in {1,...,4}
            \path[yshift=-1.0cm]
                node[hidden neuron] (H-\name) at (\layersep,-\y cm) {};
    
        \path[yshift=-2.0cm] node[latent neuron] (L-1) at (5.0cm, -1cm) {};
        \path[yshift=-2.0cm] node[latent neuron, label=below:$z$] (L-2) at (5.0cm, -2cm) {};
    
        \foreach \name / \y in {1,...,4}
            \path[yshift=-1.0cm]
                node[hidden neuron] (H2-\name) at (7.5cm,-\y cm) {};
    
        \foreach \name / \y in {1,...,6}
            \node[output neuron] (O-\name) at (10cm, -\y cm) {};
    
        \foreach \source in {1,...,6}
            \foreach \dest in {1,...,4}
                \path (I-\source) edge (H-\dest);
    
        \foreach \source in {1,...,4}
            \foreach \dest in {1,...,2}
                \path (H-\source) edge (L-\dest);
    
        \foreach \source in {1,...,2}
            \foreach \dest in {1,...,4}
                \path (L-\source) edge (H2-\dest);
    
        \foreach \source in {1,...,4}
            \foreach \dest in {1,...,6}
                \path (H2-\source) edge (O-\dest);
    
        \node[annot,above of=H-1, node distance=2cm] (hl) {\textbf{Hidden layer 1}};
        \node[annot,left of=hl] {\textbf{Input layer}};
        \node[annot,right of=hl] (hla) {\textbf{Latent layer}};
        \node[annot,right of=hla] (hl2) {\textbf{Hidden layer 2}};
        \node[annot,right of=hl2] {\textbf{Output layer}};

        \draw [decorate,decoration={brace,amplitude=10pt, mirror},xshift=-4pt,yshift=0pt] (-0.25cm,-0.75cm) -- (-0.25cm,-6.25cm) node [black,midway,xshift=-0.6cm] {\footnotesize $x$}; 
        \draw [decorate,decoration={brace,amplitude=10pt},xshift=-4pt,yshift=0pt] (10.5cm,-0.75cm) -- (10.5cm,-6.25cm) node [black,midway,xshift=0.6cm] {\footnotesize $y$}; 

    \end{tikzpicture}
  
  \end{center}
  \caption{A standard autoencoder.}
  \label{fig:AE}
\end{figure}

A popular candidate architecture for anomaly detection is the \textit{autoencoder} 
(AE)~\cite{bank2020autoencoders}, which has been previously studied in a particle physics context~\cite{Farina_2020, Heimel_2019}.
Autoencoders are an example of a generative model in which a network is trained to reconstruct a given input. 
Figure \ref{fig:AE} shows an example of a standard AE architecture.

A key feature of autoencoders is a latent layer in the center of the architecture which is often of a lower dimensionality than the input, directly restricting the network's ability to perfectly reconstruct its input. 
In such a case, the network achieves its training goal best when it can represent high-level features of the input as vectors, or $codes$, in its latent space. 
The accuracy with which each code represents the input can be verified by decoding it, and comparing its result with the original input. 
In this way, the AE is considered to act as two neural networks being trained in parallel: an $encoder$ network $f$
which acts as the map from data to the latent space, ${\textbf z}=f({\textbf x})$, and
a $decoder$ network $g$ which then attempts to reconstruct the original input from 
its encoded representation, ${\textbf y}=g({\textbf z})$. 
The loss function of the AE can be any function of the form $\mathcal{L}({\textbf x}, {\textbf y})=g(f({\textbf x}))$, which 
is minimal when ${\textbf y}={\textbf x}$. 
A common choice is the {\it Mean Squared Error} (MSE)
between the input and output of the autoencoder:

\begin{equation}
\mathcal{L}=|{\textbf y} - {\textbf x}|^2.
\end{equation}

In the context of anomaly detection, elements which represent a small portion of a dataset will contribute less during the training process. 
As a result, they will be less represented by the learned codes when compared to elements belonging to the majority of the data. 
One can therefore expect the reconstruction of anomalous elements to be worse, placing them in the
tails of the loss function's distribution after training. 
This principle has been explored in anomalous jet tagging, for instance by representing the jets as images~\cite{Farina_2020}, or as lists 
of constituents~\cite{Heimel_2019}.

\subsection{Variational Autoencoders}

\textit{Variational Autoencoders} (VAEs) are built on the idea of standard AEs, with the extension that they are designed to perform
Bayesian inference. 
This assumes that observed data $\textbf x$ is generated by some
hidden random variable $\textbf z$ whose posterior distribution $p(\textbf{z}|\textbf{x})$ 
is intractable. 
The goal of a VAE is to learn an approximate posterior distribution, $q(\textbf{z}|\textbf{x})$, through training.

\begin{figure}
  \begin{center}
  
    \def\layersep{2.5cm}
    
    \begin{tikzpicture}[shorten >=1pt,->,draw=black!50, node distance=\layersep]
        \tikzstyle{every pin edge}=[<-,shorten <=1pt]
        \tikzstyle{neuron}=[circle,fill=black!25,minimum size=17pt,inner sep=0pt]
        \tikzstyle{input neuron}=[neuron, fill=black!20!green];
        \tikzstyle{output neuron}=[neuron, fill=red!50];
        \tikzstyle{latent neuron musig}=[neuron, fill=black!10!yellow];
        \tikzstyle{latent neuron}=[neuron, fill=black!5!orange];
        \tikzstyle{hidden neuron}=[neuron, fill=blue!50];
        \tikzstyle{annot} = [text width=4em, text centered]
    
        \foreach \name / \y in {1,...,6}
            \node[input neuron] (I-\name) at (0,-\y) {};
    
        \foreach \name / \y in {1,...,4}
            \path[yshift=-1.0cm]
                node[hidden neuron] (H-\name) at (2.0cm,-\y cm) {};
    
        \foreach \name / \y in {1,...,2}
            \path[yshift=-1.0cm]
                node[latent neuron musig] (Lm-\name) at (4.0cm,-\y cm) {};
  
        \foreach \name / \y in {1,...,2}
            \path[yshift=-3.0cm]
                node[latent neuron musig] (Ls-\name) at (4.0cm,-\y cm) {};
  
        \foreach \name / \y in {1,...,2}
            \path[yshift=-2.0cm]
                node[latent neuron] (Lz-\name) at (6.0cm,-\y cm) {};
    
        \foreach \name / \y in {1,...,4}
            \path[yshift=-1.0cm]
                node[hidden neuron] (H2-\name) at (8cm,-\y cm) {};

        \foreach \name / \y in {1,...,6}
            \node[output neuron] (O-\name) at (10cm, -\y cm) {};
    
        \foreach \source in {1,...,6}
            \foreach \dest in {1,...,4}
                \path (I-\source) edge (H-\dest);
    
        \foreach \source in {1,...,4}
            \foreach \dest in {1,...,2}
                \path (H-\source) edge (Lm-\dest);
  
        \foreach \source in {1,...,4}
            \foreach \dest in {1,...,2}
                \path (H-\source) edge (Ls-\dest);
  
        \foreach \dest in {1,...,2}
            \path (Lm-\dest) edge (Lz-\dest);
  
        \foreach \dest in {1,...,2}
            \path (Ls-\dest) edge (Lz-\dest);
    
        \foreach \source in {1,...,2}
            \foreach \dest in {1,...,4}
                \path (Lz-\source) edge (H2-\dest);
    
        \foreach \source in {1,...,4}
            \foreach \dest in {1,...,6}
                \path (H2-\source) edge (O-\dest);
    
        \node[annot,above of=H-1, node distance=2cm] (hl) {\textbf{Hidden layer 1}};
        \node[annot,above of=I-1, node distance=1cm] {\textbf{Input layer}};
        \node[annot] (hla) at (5cm, 0cm) {\textbf{Latent layer}};
        \node[annot,above of=H2-1, node distance=2cm] (hl2) {\textbf{Hidden layer 2}};
        \node[annot,above of=O-1, node distance=1cm] {\textbf{Output layer}};
        \node[draw, rectangle, very thick, minimum width=1cm, minimum height=2cm, label=$\boldsymbol{\mu}$] (r1) at (4cm, -2.5cm) {};
        \node[draw, rectangle, very thick, minimum width=1cm, minimum height=2cm, label=below:$\boldsymbol{\sigma}$] (r1) at (4cm, -4.5cm) {};
        \node[draw, rectangle, very thick, minimum width=1cm, minimum height=2cm, label=below:{$\textbf{z}=\boldsymbol{\mu} + \boldsymbol{\sigma} \epsilon$}] (r1) at (6cm, -3.5cm) {};

        \draw [decorate,decoration={brace,amplitude=10pt, mirror},xshift=-4pt,yshift=0pt] (-0.25cm,-0.75cm) -- (-0.25cm,-6.25cm) node [black,midway,xshift=-0.6cm] {\footnotesize $\textbf{x}$}; 
        \draw [decorate,decoration={brace,amplitude=10pt},xshift=-4pt,yshift=0pt] (10.5cm,-0.75cm) -- (10.5cm,-6.25cm) node [black,midway,xshift=0.6cm] {\footnotesize $\textbf{y}$};

    \end{tikzpicture}
    \caption{A Variational Autoencoder with a Gaussian latent space parametrization.}
    \label{fig:VAE}
  \end{center}
\end{figure}
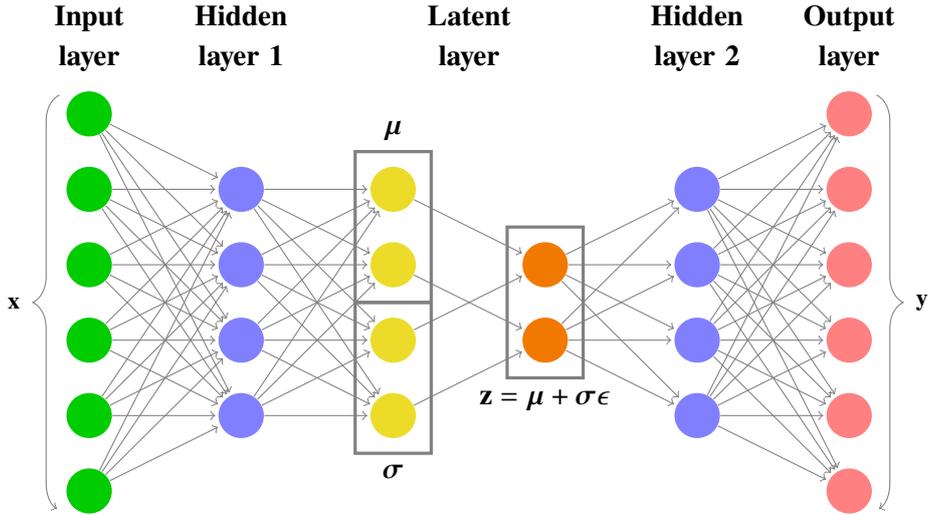

The architecture of a VAE, as shown in Figure \ref{fig:VAE}, is very close to that of a standard AE. 
The main difference is a latent space that can accommodate an encoder which maps data to a distribution in the latent space rather than a single vector. 
A common choice for the form of the latent space is a multivariate Gaussian of diagonal covariance. 
In this case, the encoder can map a given input to two independent layers, 
each with the same dimensionality as the latent space.
One of these layers represents the means of the encoded Gaussian distribution while the other represents the respective standard deviations for each dimension.
Decoding then requires sampling from this resulting distribution. 
This can be easily performed in the case of a Gaussian approximate posterior by use of the \textit{reparameterization trick}.
Instead of sampling the distribution directly, a particular value of $\textbf{z}$ can be represented in the following way:
\begin{equation}
	{\textbf z} = \boldsymbol{\mu} + \boldsymbol{\sigma} \epsilon,
\end{equation}
where $\epsilon$ is sampled from a unit isotropic normal distribution $\epsilon \sim \mathcal{N}(0, 1)$~\cite{kingma2014autoencoding}.

The VAE loss function includes both a reconstruction error term as well as an additional Kullback-Leibler (KL)-Divergence term from a chosen prior distribution $p(\textbf{z})$ to the approximate posterior distribution $q(\textbf{z}|\textbf{x})$:

\begin{equation}
	\mathcal{L} = |{\textbf y} - {\textbf x}|^{2} + D_{KL}(q({\textbf z}|{\textbf x})||p({\textbf z})).
\end{equation}

For the prior, it is common to choose a unit isotropic Gaussian centered at the origin, as the KL-Divergence from a Gaussian prior to a Gaussian approximate posterior takes on a closed form solution~\cite{Goodfellow-et-al-2016}.

Variational Autoencoders provide a number of improvements over standard Autoencoders, both as generative models~\cite{kingma2014autoencoding} and as anomaly detection tools~\cite{An2015VariationalAB}. The inclusion of a KL-Divergence term in the loss function motivates the architecture to more appropriately model unique classes of data. It also acts as another discriminatory metric, as anomalous elements are expected to have both a large reconstruction error and a large KL-Divergence when compared to nominal elements.

While VAEs have shown promise in the task of jet-level anomaly detection, they have a number of drawbacks. 
Most notably, VAEs are a fixed-length architecture, and cannot accommodate a variable number of inputs. 
When modeling jets via their constituent four-vectors, it becomes necessary to only process at most $N$ constituents, and \textit{zero-pad} the input layer when processing a jet with a number of constituents less than $N$. 
In classifier models, this is common and benign, as the loss function depends only on the output of the network and the ground truth that it is trying to reproduce. 
However, in a VAE, the input layer's neuron values are a part of its loss function (due to the MSE loss between the input and output layers).
Therefore, the zero padded elements directly correlate with the value of the loss function. 
This introduces a direct correlation between the VAE loss and the number of constituents in the input jet, which can be difficult to remove.


\section{Variational Recurrent Neural Network}

A recurrent architecture naturally circumvents this drawback since it is designed to accommodate inputs of varying length. 
In a \textit{Recurrent Neural Network} (RNN), data is input as a sequence of features. Each feature has the same fixed dimensionality, yet the sequence itself can vary in length. 
The RNN is comprised of a chain of small fixed architectures, or \textit{cells}, which expect as inputs the fixed-length feature at each element, or \textit{time-step}, in the sequence. 
While processing the sequence, the RNN updates a \textit{hidden state} at each time-step, which is carried over and accessed by the cell during the following time-step. 
The hidden state stores a long-term representation of information within the sequence, and is the key feature allowing RNNs to process sequential data of varying length. 
The RNN cell then acts as an encoder-decoder architecture which inputs the current time-step's feature and hidden state, and outputs an updated hidden state, along with an output feature if desired. 
In the interest of performing anomaly detection using a recurrent architecture, the model in this study has been chosen to be one which combines the recurrent property of RNNs with the VAE's ability to perform variational inference. 

\begin{figure}[H]
	\begin{center}
		\includegraphics[width=420pt]{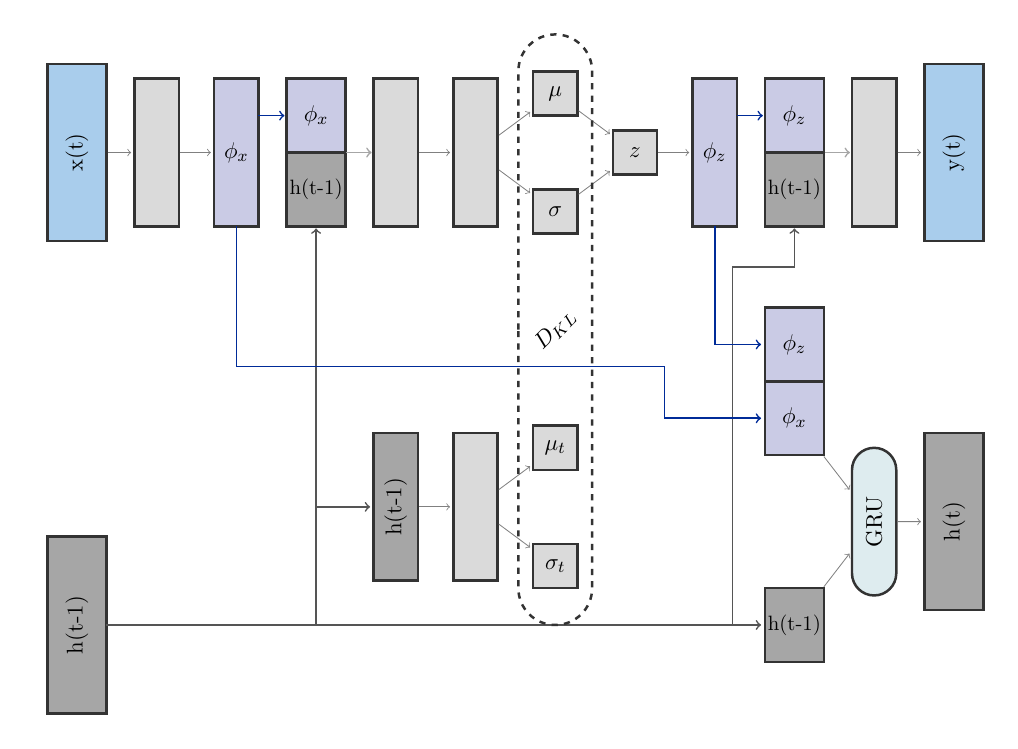}
	\end{center}
	\caption{A Variational Recurrent Neural Network cell.}
	\label{fig:VRNN}
\end{figure}

The Variational Recurrent Neural Network (VRNN) used in this study is a sequence modeling architecture
which replaces the encoder-decoder step of a traditional RNN with a VAE. 
An illustration of one VRNN cell can be seen
in Figure \ref{fig:VRNN}. 
In this model, the VAE's input at each time-step is given as the vector $x(t)$, which is then encoded and decoded into an output vector $y(t)$ which can be compared to $x(t)$ via the reconstruction loss.
The $\phi_{x}$ and $\phi_{z}$ layers represent \textit{feature-extracting layers}, which are interpreted as learned representations of the features of the input $x(t)$ and the encoded latent space distribution $z(t)$, respectively. 
After each time-step, the hidden state is updated via a recurrence relation, in which the current hidden state $h(t-1)$ and the current set of extracted features $\phi_{x}$ and $\phi_{z}$ produce an updated hidden state $h(t)$ via the following equation~\cite{chung2016recurrent}:
\begin{equation}
	h(t) = f(\phi_{x}, \phi_{z}, h(t-1)).
\end{equation} 
Performing this particular step is the primary function of traditional RNN architectures~\cite{lstm, cho2014learning}.

The VAE present in each cell of the VRNN notably differs from
conventional VAEs in the following ways:
\begin{enumerate}
  \item{The encoder and decoder are conditioned on the current time-step's hidden state.
  This is represented by the concatenation operation between the hidden state $h(t-1)$ and the feature-extraction layers $\phi_{x}$ and $\phi_{z}$.}
  \item{The prior from which the KL-Divergence is computed is no longer a unit Gaussian
  at the origin, but rather a multivariate Gaussian whose means and variances in each 
  dimension are determined from the current time-step's hidden state.}
\end{enumerate}

The inclusion of a learned, time-dependent prior distribution is an important component of the VRNN architecture. Without this feature, the decoder network would only be able to access information about the current time-step from the hidden state, and the loss function would motivate the posterior distributions for each time-step to be identical. As a result, this allows the VRNN the flexibility to model complex structured sequences with high variability, as is expected from a jet represented by a sequence of constituent four-vectors.
In more detail, each time-step's latent space prior distribution parameters $\mu_{t}$ and $\sigma_{t}$ are functions of the current time-step's hidden state~\cite{chung2016recurrent}:
\begin{equation}
	z_{t} \sim \mathcal{N}(\mu_{t}, \sigma_{t}), \text{ where } \mu_{t}, \sigma_{t} = f^{prior}(h_{t-1}).
\end{equation} 
Similarly, the latent space approximate posterior is defined by parameters $\mu$ and $\sigma$ which are functions of the input's extracted features $\phi_{x}$ and the hidden state $h_{t-1}$:
\begin{equation}
	z \sim \mathcal{N}(\mu, \sigma), \text{ where } \mu, \sigma = f^{post.}(\phi_{x}, h_{t-1}).
\end{equation} 
The generated output is then decoded from features extracted from the latent space distribution $\phi_{z} = f(z)$, while also being conditioned on the hidden state:
\begin{equation}
y(t) = f^{dec}(\phi_{z}, h(t-1)).
\end{equation} 

A loss for each time-step $\mathcal{L}(t)$ can then be computed by incorporating both the reconstruction error between the input constituent $x(t)$ and generated output constituent $y(t)$, as well as the KL-Divergence between the approximate posterior $z$ and the learned prior $z_{t}$. A constant $\lambda$ is also included which weights the KL-Divergence term's contribution to the loss:
\begin{equation}
\mathcal{L}(t) = |{\bf y}(t) - {\bf x}(t)|^{2} + \lambda D_{KL}(z||z_{t}).
\end{equation} 

An overall loss $\mathcal{L}$ over the sequence is then computed by averaging the individual time-step losses over the length of the sequence $N$:

\begin{equation}
\mathcal{L} = \frac{\mathcal{L}(t)}{N}.
\end{equation} 

This loss function performs the same role as the VAE's loss function, acting both as an appropriate means of optimizing the architecture as well as a discriminatory quantity between nominal and anomalous elements of the dataset.

The details of the VRNN architecture used in this study are as follows: the number of neurons in each intermediate layer, including the hidden state and feature extracting layers, but not including the latent space and its $\mu$ and $\sigma$ layers, is 16. The latent space is chosen to be two-dimensional. Since constituent four-vectors of jets are being modeled, the input $x(t)$ and output $y(t)$ layers are three dimensional, corresponding to the $p_{T}, \eta, $and $\phi$ of each constituent. ReLU~\cite{activations} activations are used in each layer of the network, except for $\sigma$ and $\sigma_{t}$, which have softmax~\cite{activations} activations, and $z$ and $y(t)$, which have linear activations. 

The constituents of an input jet are processed sequentially, one per time-step. 
Each time-step contributes a loss based on the VAE loss function: 
\begin{equation}
\mathcal{L}(t) = MSE + \lambda D_{KL},
\end{equation}
where $\lambda$ is a factor which weights the KL-Divergence contribution relative to the MSE reconstruction loss.

Since harder constituents contribute more information toward the identification of jet substructure, $\lambda$ is defined to be be a function of constituent $p_{T}$ fraction such that lower $p_{T}$ constituents obtain a lower weight in the loss function. 
Furthermore, since a constituent's $p_{T}$ fraction depends directly on the number of constituents in the jet, a unique $p_{T}$ fraction distribution is generated for all possible constituent multiplicities by averaging the $p_{T}$ fractions of each constituent across the entire dataset. In more detail, the dataset-averaged $p_{T}$ fraction, $\overline{p_{T}}_{N}(t)$ of the $t^{th}$ constituent in a jet, $j_{N}$, with $N \geq t$ constituents is expressed as:

\begin{equation}
\overline{p_{T}}_{N}(t)=\frac{1}{D_{N}}\displaystyle\sum_{j_{N} = 1}^{D_{N}} p_{T,j_{N}}(t),
\end{equation}

where $D_{N}$ is the number of jets in the dataset which have $N$ constituents.

The loss is therefore computed for each constituent as:
\begin{equation}
\mathcal{L}(t)=MSE+0.1\overline{p_{T}}_{N}(t)D_{KL},
\end{equation}
where $MSE$ is the mean-squared-error between $x(t)$ and $y(t)$, $D_{KL}$ is the KL-Divergence from the current time-step's prior distribution and the encoded posterior, and $\overline{p_T}_{N}(t)$ is the dataset-averaged $p_T$ fraction of the constituent at time-step $t$. The final loss is computed by averaging the individual time-step losses over the entire jet: 
\begin{equation}
\mathcal{L} = \frac{\Sigma \mathcal{L}(t)}{N}.
\end{equation} 
The hyperparameters involved in this implementation, namely the dimensionality of intermediate layers, and the additional weight coefficient of 0.1 in the loss function were determined via a hyperparameter optimization scan. 

After the network is trained, an \textit{Anomaly Score} can be determined for each jet. The KL-Divergence term has been shown to provide better discrimination between anomalous and standard jets than either the reconstruction error or the loss term as a whole. Therefore, the Anomaly Score is defined in terms of the KL-Divergence of each constituent, averaged over the whole jet, and restricted to the range of (0, 1) via exponentiation:
\begin{equation}
	\text{Anomaly Score} = 1 - e^{-\overline{D_{KL}}}.
\end{equation}


\section{Data Samples and Pre-Processing}
\label{section:data}

The performance of the VRNN is investigated by studying its ability to discriminate signal from background in a contaminated dataset of background dijet events with varying amounts of signal. The signal events are a process of the form of $Z'\rightarrow XY \rightarrow JJ$ where $X$ and $Y$ are two heavy resonances each decaying to hadrons forming a boosted jet $J$. Two types of signal events are generated. One type is comprised of events where the X and Y both decay to two quarks, resulting in boosted jets with two-pronged substructure. The other type differs in that the X and Y both decay to three quarks, resulting in boosted jets with three-pronged substructure. The masses of the particles in the signal hypothesis are 3.5 TeV, 500 GeV, and 100 GeV for $Z'$, $X$, and $Y$ respectively. A total of 99457 signal events were generated for each substructure hypothesis, along with 995,453 background events. The events were generated using {\sc Pythia8} and the detector response was simulated using {\sc Delphes 3.4.1} with no pile-up or multiple parton interactions included, and events were selected using a single large-radius ($R$=1.0) jet trigger with a $p_T$ threshold of 1.2 TeV. The dataset was provided as part of the LHC Olympics challenge for the ML4Jets2020 Workshop~\cite{dataset}. 

The data were provided as a list of hadrons for each event. The hadrons were then clustered into jets using the anti-$k_{t}$ jet-clustering algorithm with a radius parameter of 1.0~\cite{Cacciari_2008} as implemented by  {\sc FastJet 3.3.3}~\cite{fastjet}. In this study, only the highest $p_{T}$ (leading) and second-highest $p_{T}$ (sub-leading) jets are considered in each event. To test the model's performance with varying amounts of signal contamination, contaminated datasets were produced with 12 different signal event fractions, 10 of which were generated in the range of 0.01\% to 10.0\% along a logarithmic scale, with two higher signal event fractions of 25\% and 50\%. For contamination levels up to and including 10\%, contaminated datasets were created using the same set of background events while only the amount of signal was varied to match the desired contamination. For the two highest contamination levels, contaminated datasets were created using the same 99457 signal events while the number of background events was limited.
Independent datasets were generated for both two-prong and three-prong signal substructure hypotheses at each level of contamination.
The events contained in contaminated datasets are unlabeled, such that it is unknowable from which set they were sourced.

Since the goal is to identify jets mainly due to their substructure, it is important that the model's Anomaly Score does not correlate with other jet features, namely mass and $p_{T}$. A common practice to avoid such a correlation in neural network jet modeling architectures is the use of adversarial de-correlation networks~\cite{louppe2017learning}. Applying such adversarial architectures to a VRNN is a complex task which is outside of the scope of this study~\cite{Purushotham2017VariationalRA}. Instead, the de-correlation is achieved through a pre-processing procedure, which can be divided into two steps: one which directly removes mass and $p_{T}$ information from the input jets, and another which orders constituents in a way that improves the VRNN's discriminatory performance.

\subsection{Jet Alignment}

The pre-processing method is developed to produce jets which are superficially identical, with the only differences appearing in the arrangement and energy distribution of their constituents due do varying substructure. This procedure is inspired by a study based on jet images, where a pre-processing method which boosts each jet to the same reference frame allows for a model trained on the pre-processed jets to be robust against variations in mass and $p_{T}$~\cite{roy2020robust}. The process can be briefly summarized in three steps:

\begin{itemize}
	\item{Rescale each jet to the same mass,}
	\item{Boost each jet to the same energy,}
	\item{Rotate each jet to the same orientation in $\eta$, $\phi$.}
\end{itemize}

Algorithm~\ref{alg:alignment} describes in detail the implementation of the rescaling, boosting, and rotating processes, referred to as \textit{alignment} in this study. 

\begin{algorithm}[H]
\SetAlgoLined
\SetKw{Beginning}{Start}
  \Beginning{}

  Boost jet in $z$ direction until $\eta_{Jet} = 0$
 
  Rotate jet about $z$ axis until $\phi_{Jet} = 0$
 
  Rescale jet four-vector such that $m_{Jet} = $0.25 GeV
 
  Boost jet along its axis until $E_{Jet} = $1 GeV
 
  Rotate jet about $x$ axis until hardest constituent has $\eta_{1} = 0, \phi_{1} > 0$

  \eIf{Any constituents have $\Delta R > 1$\footnote{$\Delta R$ is computed as $\sqrt{\eta^2 + \phi^2}$ for each constituent, where $\eta$ and $\phi$ are measured relative to the x axis.} }{
 
  Remove all constituents with $\Delta R > 1$
 
  Rebuild jet with remaining constituents

  Repeat from start  
  
  }{
  continue
  } 
 
 \eIf{Number of constituents $>$ 20}{ 
 
 Keep up-to the first 20 constituents, ordered in $p_T$
 
 Rebuild jet with remaining constituents
 
 Repeat from start
 }{
 continue
 }

Reflect constituents about $\phi$ axis such that the second hardest constituent has $\eta_{2} > 0$
 
\caption{Jet Alignment}
\label{alg:alignment}
\end{algorithm}

\begin{figure}[H]
	\begin{center}
		\includegraphics[width=213pt]{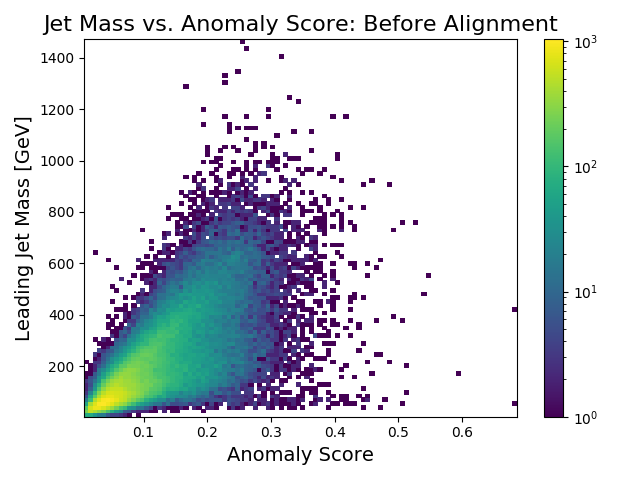}
		\includegraphics[width=213pt]{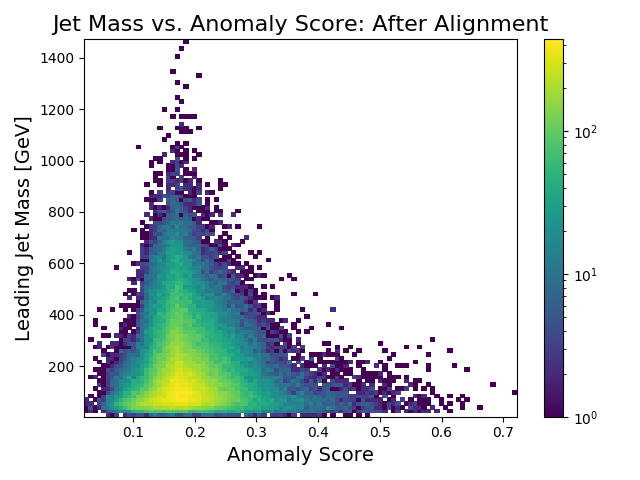}
	\end{center}
	\caption{Leading jet mass vs Anomaly Score distributions before (left) and after (right) applying the alignment method detailed in Algorithm~\ref{alg:alignment}.}
	\label{fig:mass_vs_score_boost}
\end{figure}

To evaluate the efficacy of this procedure, the model is trained and evaluated on a dataset of background jets both before and after alignment, and the resulting correlation between Anomaly Score and jet mass is compared. Figure \ref{fig:mass_vs_score_boost} shows the two-dimensional distribution of the mass of the highest $p_T$ (leading) jet in each event vs. its Anomaly Score before and after aligning the input jets. The results depict a significantly smaller amount of correlation between the jet's mass and its Anomaly Score after alignment, as desired.

\subsection{Sequence Ordering}

After the alignment step has been performed, the effect of \textit{sequence ordering} on the input constituents has additionally been investigated. In fixed architecture models, such as VAEs or image-based Convolutional Neural Networks (CNNs), the ordering of constituents in the list of training inputs is seldom important. However, in recurrent architectures such as the VRNN, choosing a sequence ordering method that highlights important sequence features can boost performance.
 
The objective of this study is to build a model which can differentiate between isotropic jets resulting from soft QCD interactions, and jets with multiple cores resulting from the hadronic decay of boosted objects. Therefore, it is favorable to use a sequence ordering which makes the existence of multiple hard cores of a jet distinctly apparent. This is achieved by ordering the constituents in $k_{t}$-distance order. More specifically, the $n^{th}$ constituent in the list is determined to be the constituent with the highest $k_{t}$-distance relative to the previous constituent, with the first constituent in the list being the highest $p_{T}$ constituent after alignment:
\begin{equation}
	c_{n} = max(p_{Tn}\Delta R_{n, n-1}).
\end{equation}

\begin{figure}[H]
	\begin{center}
		\includegraphics[width=213pt]{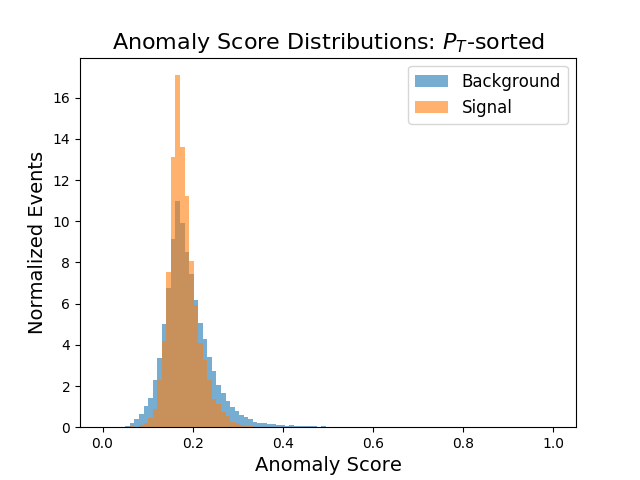}
		\includegraphics[width=213pt]{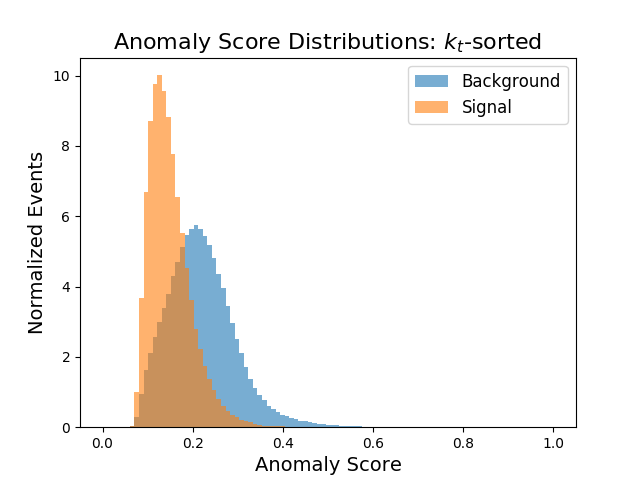}
	\end{center}
	\caption{Leading jet Anomaly Score distributions for background and two-prong signal jets after the network was trained on a 10\% contaminated training set, with $p_T$-sorted (left) and $k_{t}$-sorted (right) ordering of constituents for input jets.}
	\label{fig:score_comp}
\end{figure}

The effect on performance due to this choice of constituent ordering can be easily illustrated in the case of a two-prong jet. In such a case, the sequence will start with a constituent in one of the two cores of the jet, and be subsequently followed by a constituent belonging to the other core and so on. This results in an easily predictable pattern which the VRNN is better able to identify, particularly compared to a background jet. The resulting performance difference between $p_{T}$-sorted and $k_{t}$-sorted inputs is shown in Figure \ref{fig:score_comp}. Using a 10\% contaminated dataset as a training set, and then comparing Anomaly Score distributions determined from leading jets of background-only and two-prong signal-only datasets, the discrimination between signal and background jets is notably better when the VRNN input sequence allows for detection of multi-prong substructure early on. It is also important to note that the signal jets are assigned a lower Anomaly Score than the background jets. This can be attributed to the same reason signal and background jets are distinguishable after applying $k_{t}$-sorted sequencing: jets with multi-prong substructure are more easily modeled by the VRNN, and therefore result in an overall lower loss value when compared to background jets. 

\section{Results}

One way in which the VRNN's performance is studied is by assessing signal acceptance and background rejection at the jet level by using only the leading jet of each event. In addition, the Anomaly Score can be applied to both the X and Y jets in an event and used to discriminate between signal and background in an event-level analysis context. Results of the VRNN's performance are provided for both approaches below.

Training is performed using the {\sc PyTorch} deep learning library~\cite{pytorch}. The network is updated using the Adam optimizer with a learning rate parameter of $10^{-5}$~\cite{kingma2017adam}. No regularization via weight decay is applied, however gradient clipping is implemented with a clip value of 10. Since the training scenario is entirely unsupervised, the resulting Anomaly Score distributions from each training dataset may vary. To arrive at a consistent score distribution, a transformation is applied on the resulting Anomaly Score which aims to satisfy two conditions:

\begin{itemize}
	\item{The mean of the resulting distribution is at an Anomaly Score value of 0.5.}
	\item{Anomaly Scores closer to a value of 1 correspond to more signal-like jets. Note that this reverses the previously observed feature displayed in Figure~\ref{fig:score_comp} where more signal-like jets are assigned a lower Anomaly Score.}
\end{itemize}

The transformation can be summarized as
\begin{equation}
\label{eq:transformation}
	\rho ' = 1 - \bigg(\frac{\rho}{2\overline{\rho}}\bigg),
\end{equation}

where $\rho '$ is the transformed Anomaly Score, and  $\overline{\rho}$ is the mean of the un-transformed Anomaly Score distribution of the training set.

\subsection{Jet Level Performance}

\begin{figure}[H]
	\begin{center}
		\includegraphics[width=250pt]{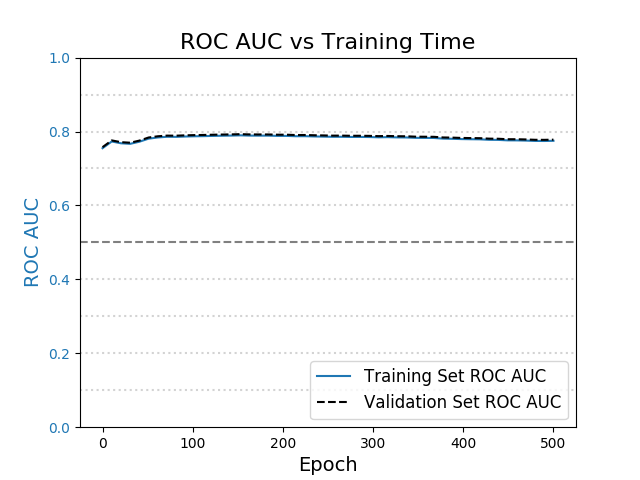}
	\end{center}
	\caption{Area Under the Curve (ROC AUC) vs. training time in epochs on a 1\% signal-contaminated dataset. The VRNN reaches an optimal performance quickly, and retains this performance over a long training period. The difference in performance between the training and validation sets is a result of the former containing elements of signal.}
	\label{fig:auc_vs_epoch}
\end{figure}

In the jet-level assessment, the model is trained on the leading jet of each event for 500 epochs. To evaluate the trend in performance during training, a computation of the Receiver Operating Characteristic's Area Under the Curve (ROC AUC) is performed after each epoch by examining events in the contaminated training set or comparing events in the background-only validation set to those in the signal-only set. Figure \ref{fig:auc_vs_epoch} shows the results of this training scenario in the case of 1\% contamination. The VRNN quickly reaches its optimal performance, and retains a stable performance throughout the training period. 

\begin{figure}[H]
	\begin{center}
		\includegraphics[width=213pt]{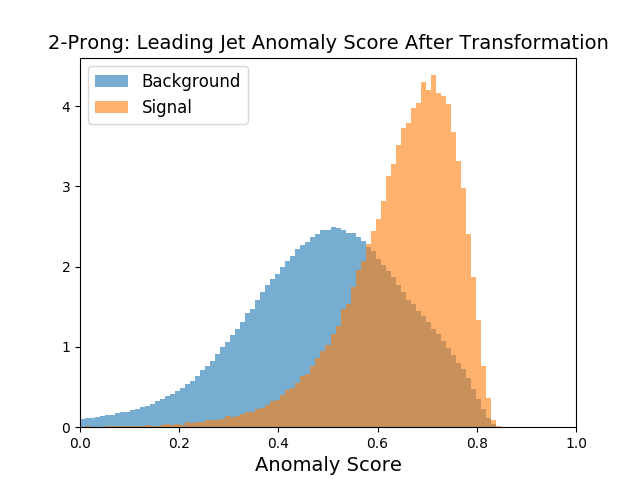}
		\includegraphics[width=213pt]{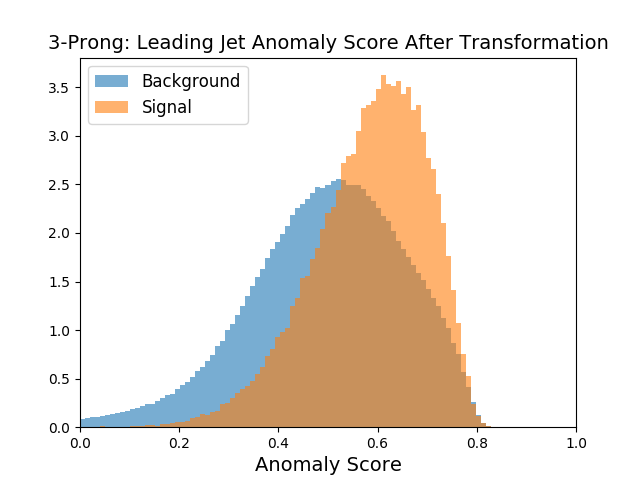}
	\end{center}

	\caption{Anomaly Score distributions from training over a dataset with 10\% signal contamination after applying the transformation described in Equation (\ref{eq:transformation}). The Anomaly Score in these figures is computed from the leading jets of each event, for both the background sample and two-prong (left) or three-prong (right) signal samples.}

	\label{fig:score_transform}
\end{figure}

To evaluate the model's performance, the weights corresponding to a training period of 100 epochs were chosen in the following studies. Figure \ref{fig:score_transform} shows the distributions of the Anomaly Score for leading jets in the background sample and both the two-prong and three-prong signal samples after training over a contaminated training set with 10\% signal contamination and applying the transformation in Equation (\ref{eq:transformation}).

\begin{figure}[H]
	\begin{center}
		\includegraphics[width=213pt]{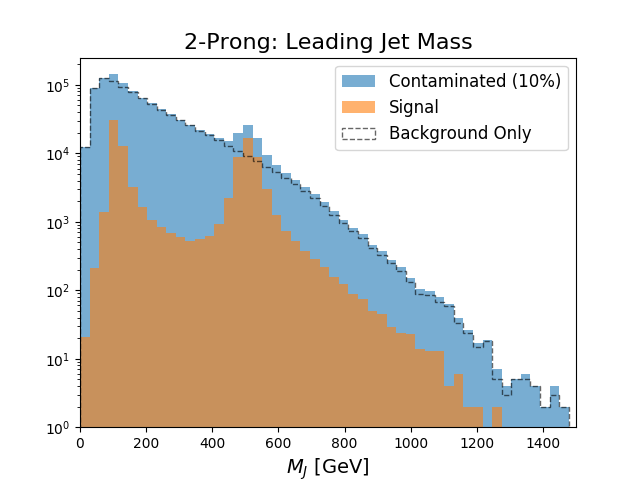}
		\includegraphics[width=213pt]{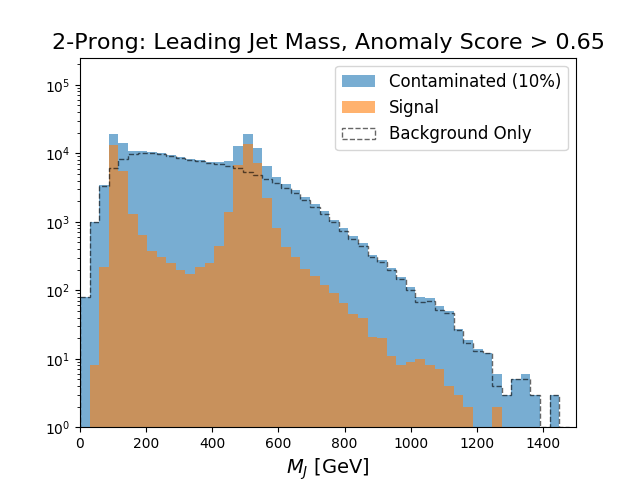}
	\end{center}
	\caption{Leading jet mass distributions with a two-prong signal hypothesis before (left) and after (right) requiring the Anomaly Score to exceed a value of 0.65.}
	\label{fig:2P_lj_mass}
\end{figure}

\begin{figure}[H]
	\begin{center}
		\includegraphics[width=213pt]{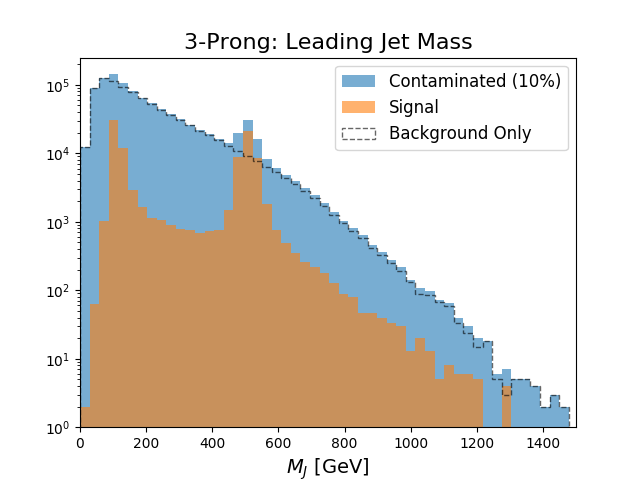}
		\includegraphics[width=213pt]{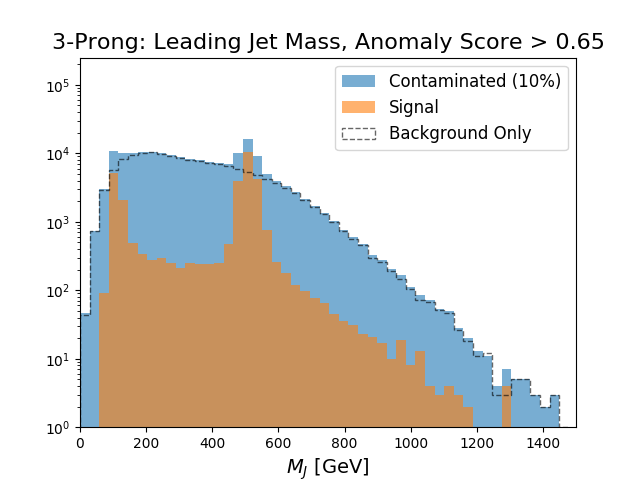}
	\end{center}
	\caption{Leading jet mass distributions with a three-prong signal hypothesis before (left) and after (right) requiring the Anomaly Score to exceed a value of 0.65.}
	\label{fig:3P_lj_mass}
\end{figure}

Figure \ref{fig:2P_lj_mass} shows the mass distributions of the leading jet in signal, background-only, and signal-contaminated datasets, before and after a jet-level selection requiring the Anomaly Score to exceed a value of 0.65, corresponding to a rejection factor of about 6 for background jets.
This value is chosen in the interest of displaying the discriminating power of the Anomaly Score while retaining enough background statistics to observe the background shape sculpting. 
The visibility of the resonances at 100 GeV and 500 GeV is enhanced after the selection. Sculpting in the background distribution is observed, which is an effect of mass correlation mainly introduced by the $k_{t}$-ordered sequencing, as there is a correlation between the number of hard cores in a jet and its mass. However, the observed sculpting is mainly a suppression of low mass events, and does not result in the generation of peaks in the mass distribution. Both the signal enhancement and background sculpting are similarly observed on three-pronged signatures in Figure \ref{fig:3P_lj_mass}, also shown for a 10\% contaminated dataset and an Anomaly Score selection of greater than 0.65.  

As the Anomaly Score in this context distinguishes multi-pronged substructure from homogenous jets, it is useful to compare it to a commonly used high-level variable sensitive to two-pronged signals. The energy correlation function ratio $D_2$~\cite{d2} is selected and used as a benchmark to contextualize the Anomaly Score in both signal discrimination and jet mass correlation. The $D_2$ variable is such that values closer to zero correlate with two-prong jet substructure. 

\begin{figure}[H]
	\begin{center}
		\includegraphics[width=250pt]{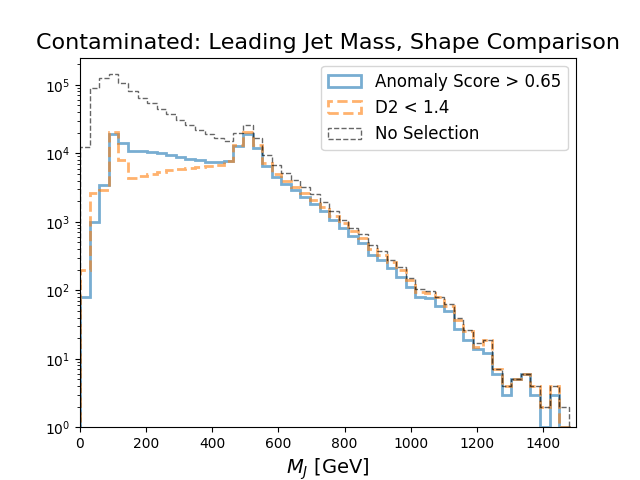}
	\end{center}
	\caption{Comparison of the leading jet mass distribution in a contaminated dataset between equivalent background acceptance selections on Anomaly Score and the $D_2$ variable. The $D_2$ selection causes more severe sculpting in the jet mass distribution than the Anomaly Score, indicating that selections on the Anomaly Score provide a more faithful representation of the original background mass distribution while still enhancing the presence of signal-like jets.}
	\label{fig:d2_comp}
\end{figure}

Figure \ref{fig:d2_comp} shows a comparison of the shapes of the contaminated jet mass distributions with a two-prong signal hypothesis when subject to a selection on Anomaly Score or $D_2$. The Anomaly Score is again selected to exceed a value of 0.65, and the $D_2$ selection is chosen to be less than 1.4 to provide an equivalent background rejection factor of about 6. The shape of the jet mass distribution is more significantly sculpted after the $D_2$ selection than the Anomaly Score selection, as can be seen most prominently in the 200 to 400 GeV range of jet mass. This indicates that there is a more significant correlation of $D_2$ with jet mass while the Anomaly Score selection retains more of the smoothly falling characteristics of the background jet mass distribution. Such a result can be attributed largely to the alignment method used during pre-processing, as well as to the Anomaly Score being determined only from jet constituent four-vector information, without any high-level information being input into the model. 

\begin{figure}[H]
	\begin{center}
		\includegraphics[width=250pt]{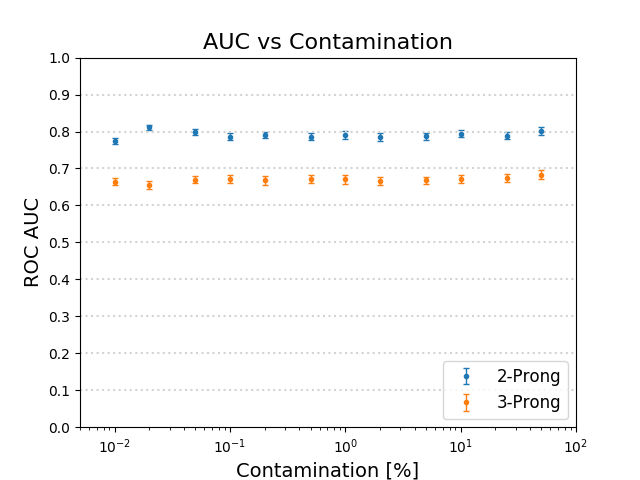}
	\end{center}
	\caption{ROC AUC vs. percent signal contamination in training datasets. The performance of the Anomaly Score is consistent across a wide range of contamination levels.}
	\label{fig:aucs_vs_contam}
\end{figure}

Another important study involves the model's performance over a range of signal contamination
levels. Figure~\ref{fig:aucs_vs_contam} shows the ROC AUC values of both two and three-pronged signal hypotheses
after training on each of the contaminated datasets described in Section 3. 
At each level of contamination, the VRNN is trained on the leading jets of both the respective two-prong and three-prong contaminated datasets for 100 epochs. 

The resulting trained network is then used to assign an Anomaly Score to each leading jet in the dataset. 
AUC values for each level of contamination are determined from a ROC curve built from 1000 randomly selected jets from both the background and signal sets after training.
Error bars are computed by repeating this process 100 times and determining the standard deviation of the resulting distribution of AUC values.
Notably, the performance is consistent along all contaminations, and able to distinguish both two and three-pronged signals without any prior substructure hypothesis. 
The Anomaly Score can therefore be interpreted as a quantity which is capable of adequately and consistently parametrizing multiple distinct substructure scenarios. 
This feature is valuable in model-independent searches, or those without a pre-defined signal substructure hypothesis. 

The ability of the Anomaly Score to be consistently performant along a large range of contaminations is unexpected in the context of anomaly detection, where the dilution of the training set with a high number of signal elements is expected to result in lower performance.
The consistent performance observed can be attributed to the choice of $k_t$-ordered sequencing and the representation of jets as variable-length sequences of constituents. Since the choice of $k_t$-ordered sequencing highlights the presence of multiple hard cores within a jet, the VRNN's Anomaly Score is predisposed to correlate with signal jets due to their anomalous substructure regardless of the level of contamination.

\subsection{Event Level Performance}

A natural benchmark of the Anomaly Score's ability to distinguish anomalous jets is to apply the score in an analysis-like context. In this study, the goal is to reconstruct the $Z'$ particle in the invariant mass spectrum $M_{JJ}$ of the two boosted jets corresponding to the decays of X and Y as described in Section \ref{section:data}. To do this, the network is trained on both the leading and sub-leading jets, with one set of network weights saved for each amount of contamination. 

\begin{figure}[H]
	\begin{center}
		\includegraphics[width=250pt]{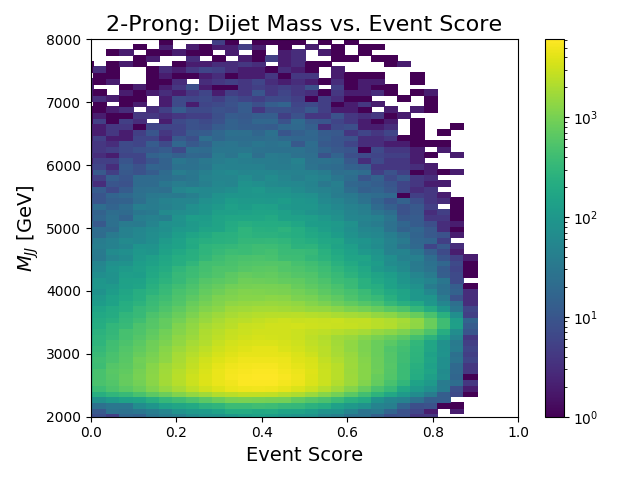}
	\end{center}
	\caption{Dijet invariant mass vs. Event Score. The presence of the resonant 3500 GeV Z' particle is visible due to events which have a high value of the Event Score.}
	\label{fig:mjj_vs_evscore}
\end{figure}

Since the model produces one Anomaly Score per jet, the Anomaly Scores for the leading and sub-leading jet must be combined to arrive at an overall \textit{Event Score}. In this study, the Event Score is chosen to be the highest of the two Anomaly Scores between the leading and sub-leading jets, resulting in an event-level discriminant which uses the most anomalous jet in the event to discriminate. The ability of the Event Score to distinguish signal from background is illustrated in Figure \ref{fig:mjj_vs_evscore}, showing the correlations between the dijet invariant mass and the Event Score in a dataset with 10\% signal contamination. The significant feature of the $3500$ GeV $Z'$ occupies a region of high Event Score, validating this quantity as a discriminant of anomalous events from background.

The goal of this search is to observe an excess of signal events in the dijet invariant mass distribution at a mass corresponding to the Z' particle. 
This is commonly referred to as a "bump hunt" search, in which the signal is expected to appear as a bump upon an otherwise smoothly falling background distribution.\footnote{This model was also demonstrated in a similar context in the LHC Olympics Anomaly Detection challenge as part of the ML4Jets2020 workshop~\cite{kasieczka2021lhc}.}

\begin{figure}[H]
	\begin{center}
		\includegraphics[width=213pt]{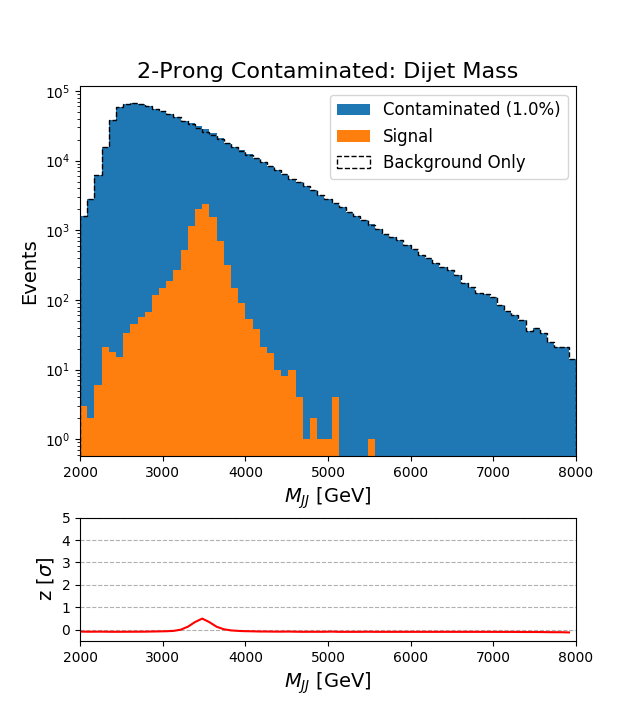}
		\includegraphics[width=213pt]{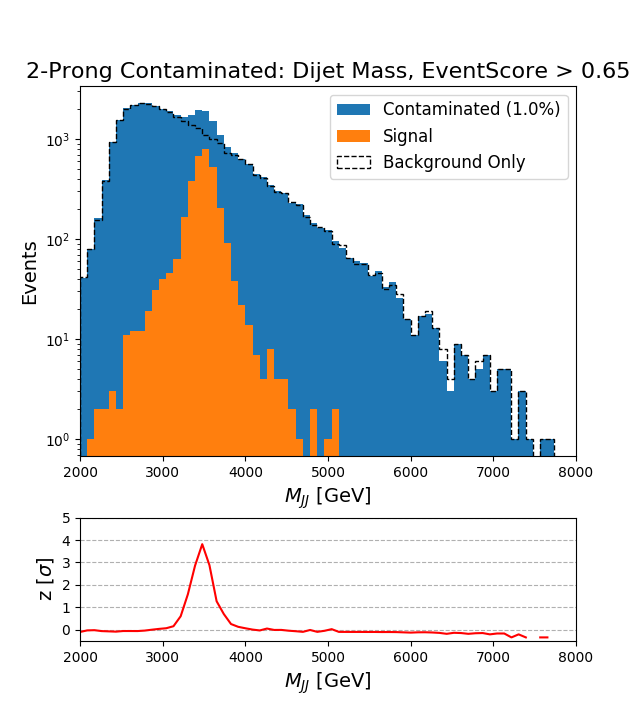}
	\end{center}
	\caption{Two-prong dijet mass distributions before (left) and after (right) requiring the Event Score to exceed a value of 0.65, at a signal contamination of 1.0\%. The Event Score selection provides an improvement in signal sensitivity from $0.5\sigma$ to $4\sigma$ while retaining the smoothly falling background distribution.}
	\label{fig:2p_dijet}
\end{figure}

Figure \ref{fig:2p_dijet} shows the dijet mass distributions of the signal, background, and contaminated datasets, before and after applying a selection on the Event Score at a value of 0.65 (the same value as was used in the jet-level performance assessment) providing a background rejection factor of about 33. Also plotted is the local significance $z$ in each bin of the corresponding histogram, where a total uncertainty of 15\% on the number of background events is assumed. The local significance is computed using the {\sc BinomialExpZ} function from {\sc RooStats}~\cite{moneta2011roostats}. 

\begin{figure}[H]
	\begin{center}
		\includegraphics[width=213pt]{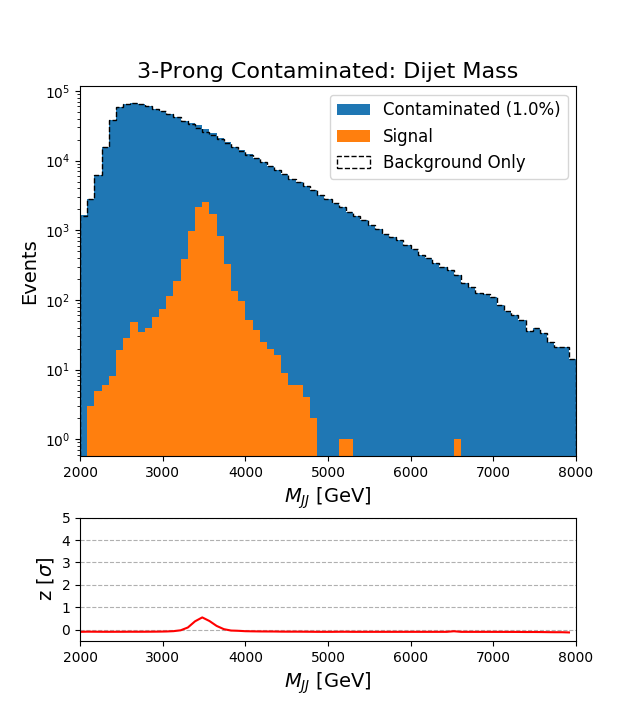}
		\includegraphics[width=213pt]{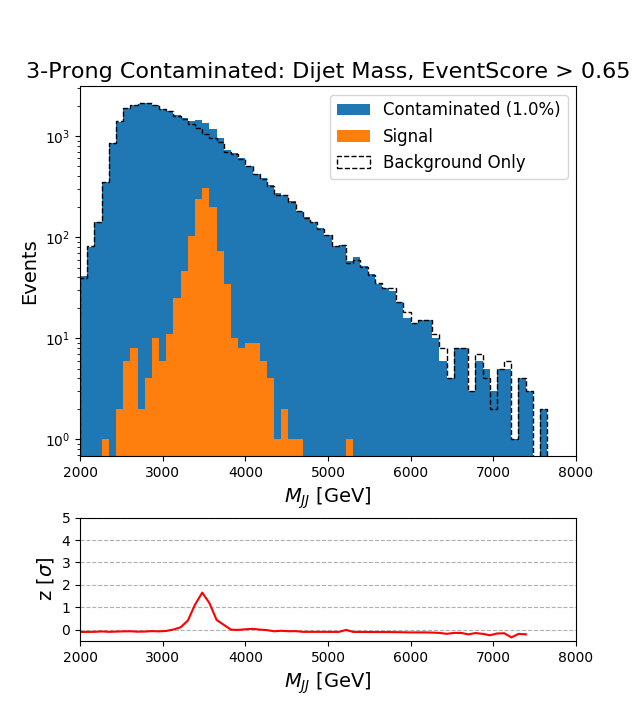}
	\end{center}
	\caption{Three-prong dijet mass distributions before (left) and after (right) requiring the Event Score to exceed a value of 0.65, at a signal contamination of 1.0\%. The Event Score selection provides an improvement in signal sensitivity from $0.5\sigma$ to $1.5\sigma$ while retaining the smoothly falling background distribution.}
	\label{fig:3p_dijet}
\end{figure}

The selection on the Event Score increases the significance of the excess from $0.5\sigma$ to $4\sigma$ at a signal contamination of 1.0\%, while still retaining the smoothly falling behavior of the background. No selections other than the Event Score requirement have been applied in these scenarios besides the initial trigger requirement of $p_{T} > 1.2$ TeV on the leading jet. In Figure \ref{fig:3p_dijet}, a similar result is seen in the case of the three-pronged signal, where the Event Score requirement results in an increase of local significance of the signal peak from $0.5\sigma$ to $1.5\sigma$ at a signal contamination of 1\% under the same conditions. These results display the capability of the Anomaly Score as an analysis variable, as it can distinguish signal events with multiple substructure hypotheses while being robust against ambiguities in signal jet mass and $p_{T}$.

\clearpage

\section{Conclusion}

A novel approach for unsupervised signal identification in the context of new physics searches is presented. 
The technique utilizes a Variational Recurrent Neural Network trained on contaminated datasets to distinguish jets resulting from boosted hadronically decaying objects from those resulting from soft QCD processes. 
A pre-processing procedure (Jet Alignment) is developed, in which each input jet is boosted to the same reference mass, energy, and orientation, coupled with a sequence ordering which makes the presence of signal-like substructure more apparent. 
The resulting training produces an Anomaly Score per jet that is sensitive to multiple substructure hypotheses. 
In a jet-level study that only uses the leading jet in each event, the model enhances signal with less jet mass correlation than the substructure-based variable $D_2$. 
In addition, the resulting Anomaly Score is equally performant across varying levels of contamination, allowing for a consistent characterization of substructure regardless of the amount of signal present in the dataset. 
When applied to an event-level context, a selection on the maximum of the two leading jet Anomaly Scores increases the significance of both two- and three-prong signals, while mostly retaining the smoothly falling shape of the background's mass distribution.

The Variational Recurrent Neural Network used in this study is a powerful tool capable of learning underlying features of physics objects presented as sequential data. 
Its applications to new physics searches are numerous, with one of the most attractive features being the potential for training directly on data without a pre-defined signal substructure hypothesis. 
It is also a general tool for modeling sequential data of any type, making it compatible with common high energy physics tasks such as event-level searches or object-level classification.

Since the overall structure of the model contains both elements of Variational Autoencoders and Recurrent Neural Networks, more complicated architectural iterations can be employed as natural extensions of the VRNN. 
Examples of possible additions include adversarial mass de-correlation networks, and conditional architectures which can supplement the VRNN's input by a fixed length vector of high-level features. Other potential studies include further investigation of the ordering of the input constituent sequence, which may be tuned to accommodate better defined model hypotheses, or a study of multi-class identification in which two or more types of anomalies are identified within the same dataset.


\clearpage

\section*{Acknowledgements}

The authors would like to thank Gregor Kasieczka, Ben Nachman, and David Shih, the organizers of the LHC Olympics 2020 Anomaly Detection Challenge, for providing the datasets used in this study and for the opportunity to develop and test the VRNN architecture.

This material is based upon work supported by the National Science Foundation under Grant No. PHY-2013070.

IO is supported by the fellowship LCF/BQ/PI20/11760025 from ”la Caixa” Foundation (ID 100010434) and by the European Union’s Horizon 2020 research and innovation programme under the Marie Skłodowska-Curie grant agreement No 847648.


\bibliographystyle{jhep}
\bibliography{vrnn}
\end{document}